# Information structure and general characterization of Mueller matrices

José J. Gil,[1,*] Ignacio San José.[2]

[1]*Departamento de Física Aplicada, Universidad de Zaragoza, Pedro Cerbuna 12, 50009 Zaragoza, Spain*
[2]*Instituto Aragonés de Estadística, Gobierno de Aragón, Bernardino Ramazzini 5, 50015 Zaragoza, Spain*

\* ppgil@unizar.es

Linear polarimetric transformations of light polarization states by the action of material media are fully characterized by the corresponding Mueller matrices, which contain in an implicit and intricate manner all measurable information on such transformations. The general characterization of Mueller matrices relies on the nonnegativity of the associated coherency matrix, which can be mathematically formulated through the nonnegativity of its eigenvalues. The enormously involved explicit algebraic form of such formulation prevents its interpretation in terms of simple physical conditions. In this work, a general and simple characterization of Mueller matrices is presented based on their statistical structure. The concepts associated with the retardance, enpolarization and depolarization properties as well as the essential coupling between the two latter are directly described in the light of the new approach.

## I. INTRODUCTION

Mueller matrices provide complete measurable (phenomenological) information on linear polarimetric interactions. Consequently, Mueller polarimetry involves different experimental and theoretical powerful techniques that make it very widely used and exploited in a great variety of applications in science, engineering, medicine, remote sensing, etc. Nevertheless, Mueller matrices have an intricate mathematical structure that makes it difficult to interpret them and transform its sixteen elements into meaningful physical descriptors. Thus, the advances in the understanding and interpretation of the physical information provided by measured Mueller matrices constitute today an important objective.

The contributions of many authors have led to a deep knowledge of the nature and characteristic properties of Mueller matrices [1-17]; nevertheless, certain aspects still remain unclear, as for instance the origin of the essential coupling between enpolarizing and depolarizing effects encoded in a Mueller matrix. Furthermore, the mathematical characterizations of Mueller matrices via the nonnegativity of their associated coherency matrices [2,10,16] or through the normal form [4,5-9,12,14], have no direct and simple relation to the fundamental intrinsic quantities like the polarizance, the diattenuation and the degree of spherical purity [18].

In this work, through the analysis of the statistical structure of Mueller matrices, two objectives are achieved, namely (1) the general characterization of Mueller matrices in terms of very simple, meaningful and explicit mathematical expressions, and (2) the physical interpretation of the information provided by a Mueller matrix in terms of objective and relevant quantities.

The approach presented is based on two main steps, namely, the transformation of a Mueller matrix to its associated arrow form, and the parameterization of the associated coherency matrix, based on decoupled statistical parameters.

The present work is organized as follows; the necessary background concepts for the description of the new results are summarized in Section II; the intrinsic standard deviation and intrinsic correlation matrices associated with a Mueller matrix **M** are defined and described in Section III; the number of arbitrary components of **M** is studied in Section IV in terms of the rank of the intrinsic correlation matrix; the correlation Mueller matrix, which encompasses, in a decoupled manner, the information of the enpolarizing properties of the interaction represented by **M** is introduced in Section V; a simple and meaningful general characterization of Mueller matrices is stated in Section VI; the structure and significance of the polarimetric information supported by **M** are analyzed in Section VII; the characterization of both nonenpolarizing and symmetric Mueller matrices is performed and briefly discussed in Sections VIII and IX, respectively, and section X is devoted to the conclusion.

## II. THEORETICAL BACKGROUND

The concept of Mueller matrix relies on that of Jones matrix. In fact, any polarimetric linear interaction, where the Stokes vector **s** of the incident electromagnetic wave is transformed to another Stokes vector $\mathbf{s}' = \mathbf{M}\mathbf{s}$, **M** being a Mueller matrix, can be considered as an ensemble average of basic interactions that can be represented by respective Jones matrices, in such a manner that **M** can be expressed as [17,19]

$$\mathbf{M} = \mathcal{L}\langle \mathbf{T} \otimes \mathbf{T}^* \rangle \mathcal{L}^\dagger, \quad \mathcal{L} \equiv \frac{1}{\sqrt{2}}\begin{pmatrix} 1 & 0 & 0 & 1 \\ 1 & 0 & 0 & -1 \\ 0 & 1 & 1 & 0 \\ 0 & i & -i & 0 \end{pmatrix}, \quad (1)$$

where superscripts * and † stand for complex conjugate and conjugate transpose, respectively, the brackets indicate ensemble average, and the 2×2 complex matrix **T**, called the Jones generator [17], in general fluctuates as a consequence of the spatial or spectral partial coherence of the nature of the interaction phenomenon [20-23]. That is, even though the interaction of a photon with a single atom or molecule is necessarily nondepolarizing and therefore can be represented through the Jones formalism, the overall macroscopic interaction together with its measurement (which involves a measurement time usually much larger than the *polarization time* [24-26] of the emerging polarization state) results in the average indicated in Eq. (1). In words of Parke III (whose PhD supervisor was Prof. Hans Mueller) "a proper statistical average of deterministic Jones devices bears the same relation to the phenomenological Mueller approach that statistical mechanics bears to thermodynamics" [19].

The statistical nature of the structure of a Mueller matrix becomes evident when, through the expansion of Eq. (1) its elements $m_{ij}$ ($i,j=0,1,2,3$) are expressed in terms of combinations of second-order moments of the fluctuating elements $t_{kl}$ ($k,l=1,2$) of $\mathbf{T}$, so that $m_{ij}$ can be combined and rearranged into a Hermitian matrix $\mathbf{H}$ that has the mathematical structure of a covariance matrix [2,10,16], which therefore is necessarily positive semidefinite. Given $\mathbf{H}$, any matrix of the form $\mathbf{UHU}^{\dagger}$, $\mathbf{U}$ being a unitary matrix, has also the formal structure of a covariance matrix and has a biunivocal relation to $\mathbf{M}$. In particular, the so-called coherency matrix $\mathbf{C}$ [2], whose elements $c_{ij}$ are linked to $m_{ij}$ by [23]

$$c_{ij} = \frac{1}{4}\mathrm{tr}\left(\mathbf{G}_i^*\mathbf{G}_j\mathbf{M}\right), \quad m_{ij} = \mathrm{tr}\left(\mathbf{G}_i^*\mathbf{G}_j\mathbf{C}\right),$$

$$\mathbf{G}_0 \equiv \begin{pmatrix} 1 & 0 & 0 & 0 \\ 0 & 1 & 0 & 0 \\ 0 & 0 & 1 & 0 \\ 0 & 0 & 0 & 1 \end{pmatrix}, \quad \mathbf{G}_1 \equiv \begin{pmatrix} 0 & 1 & 0 & 0 \\ 1 & 0 & 0 & 0 \\ 0 & 0 & 0 & i \\ 0 & 0 & -i & 0 \end{pmatrix}, \quad (2)$$

$$\mathbf{G}_2 \equiv \begin{pmatrix} 0 & 0 & 1 & 0 \\ 0 & 0 & 0 & -i \\ 1 & 0 & 0 & 0 \\ 0 & i & 0 & 0 \end{pmatrix}, \quad \mathbf{G}_3 \equiv \begin{pmatrix} 0 & 0 & 0 & 1 \\ 0 & 0 & i & 0 \\ 0 & -i & 0 & 0 \\ 1 & 0 & 0 & 0 \end{pmatrix},$$

has the peculiarity that its diagonal elements only depend on the diagonal elements of $\mathbf{M}$ and vice versa, which simplifies certain important relations that will be studied

Note that, even though $\mathbf{C}$ is defined from the second-order moments $\langle t_{kl} t_{mn}^* \rangle$ ($k,l,m,n=1,2$) of the elements of the Jones generator, its elements are not directly $\langle t_{ij} t_{kl}^* \rangle$, but are given by sums and differences of different groups of four of them. This fact does not prevent that the mathematical structure of $\mathbf{C}$ is formally indistinguishable from that of a covariance matrix.

For certain developments, it is convenient the use of the partitioned expression of $\mathbf{M}$ [27,28]

$$\mathbf{M} \equiv m_{00}\hat{\mathbf{M}}, \quad \hat{\mathbf{M}} \equiv \begin{pmatrix} 1 & \mathbf{D}^T \\ \mathbf{P} & \mathbf{m} \end{pmatrix} \quad (3)$$

where $m_{00}$ is the mean intensity coefficient (MIC; i.e., the transmittance or reflectance of incident unpolarized light), $\mathbf{D}$ and $\mathbf{P}$ are the diattenuation and polarizance vectors, and $\mathbf{m}$ is the corresponding 3×3 submatrix.

The fact that $\mathbf{C}$ is positive semidefinite leads to a general characterization of Mueller matrices through the nonnegativity of the four eigenvalues $(\lambda_0,\lambda_1,\lambda_2,\lambda_3)$ (covariance conditions) or other equivalent formulations [2,10,16]. In addition, the fact that polarimetric interactions cannot amplify the intensity of light, leads to an additional passivity condition $m_{00}(1+Q) \leq 1$, where $Q \equiv \max(D,P)$ [10,29]. Thus, a given 4×4 real matrix is formally a Mueller matrix if and only if it satisfies the four covariance conditions together with the passivity condition,

To introduce the arrow form $\mathbf{M}_A$ of a given Mueller matrix $\mathbf{M}$, let us first consider the transformation $\mathbf{m} = \mathbf{m}_{RO}\mathbf{m}_A\mathbf{m}_{RI}$ of the submatrix $\mathbf{m}$ of $\mathbf{M}$, where $\mathbf{m}_{RI}$ and $\mathbf{m}_{RO}$ are proper orthogonal matrices, and $\mathbf{m}_A$ is the diagonal matrix $\mathbf{m}_A = \mathrm{diag}(a_1,a_2,\varepsilon a_3)$, $a_1,a_2,a_3$ being the (nonnegative) singular values of $\mathbf{m}_A$ taken with the convention $a_1 \geq a_2 \geq a_3$ (without loss of generality), while $\varepsilon \equiv \det\mathbf{m}/|\det\mathbf{m}|$. Thus $\mathbf{M}_A$ is defined as [30,31]

$$\mathbf{M}_A \equiv m_{00}\begin{pmatrix} 1 & \mathbf{D}_A^T \\ \mathbf{P}_A & \mathbf{m}_A \end{pmatrix} = \mathbf{M}_{RO}^T \mathbf{M} \mathbf{M}_{RI}^T$$

$$\left[\mathbf{M}_{RI} = \begin{pmatrix} 1 & \mathbf{0}^T \\ \mathbf{0} & \mathbf{m}_{RI} \end{pmatrix}, \quad \mathbf{M}_{RO} = \begin{pmatrix} 1 & \mathbf{0}^T \\ \mathbf{0} & \mathbf{m}_{RO} \end{pmatrix}, \right. \quad (4)$$

$$\left. \det\mathbf{M}_{RI} = \det\mathbf{M}_{RO} = +1 \right]$$

The diattenuation and polarizance vectors of $\mathbf{M}$ are recovered from those of $\mathbf{M}_A$ through the respective transformations $\mathbf{D} = \mathbf{m}_{RI}^T\mathbf{D}_A$ and $\mathbf{P} = \mathbf{m}_{RO}\mathbf{P}_A$, where the 3×3 orthogonal matrices $\mathbf{m}_{RI}$ and $\mathbf{m}_{RO}$ are directly determined from the entrance and exit retarders $\mathbf{M}_{RI}$ and $\mathbf{M}_{RO}$ of $\mathbf{M}$ [30,31]. Recall that transformations of $\mathbf{M}$ performed through the product by Mueller matrices of retarders $\mathbf{M}_R$ (like $\mathbf{M}_{RI}$ and $\mathbf{M}_{RO}$) have the peculiarity of being reversible in the sense that do not affect to the ability to produce changes in both intensity and degree of polarization on the interacting light [31-33].

Some intrinsic quantities of $\mathbf{M}$, that are invariant under the above mentioned reversible transformations are the MIC $m_{00}$, the diattenuation $D \equiv |\mathbf{D}|$, the polarizance $P \equiv |\mathbf{P}|$, the degree of spherical purity $P_S \equiv \|\mathbf{m}\|_F/\sqrt{3}$ (where $\|\mathbf{m}\|_F$ stands for the Frobenius norm of $\mathbf{m}$) [18], the indices of polarimetric purity (IPP) $P_1,P_2,P_3$ [34], and the degree of polarimetric purity $P_\Delta$ (also called the depolarization index) [35]. For the definitions of the indicated descriptors we refer the reader to Ref. [36].

Mueller matrices which do not decrease the degree of polarization of any totally polarized incident electromagnetic wave are called *pure* (or *nondepolarizing*), and otherwise are termed *depolarizing*. Pure Mueller matrices have the genuine property that $P_\Delta = 1$, while depolarizing Mueller matrices necessarily satisfy $P_\Delta < 1$.

### III. STATISTICAL PARAMETERIZATION OF THE ARROW FORM OF A MUELLER MATRIX

Since the transformation from $\mathbf{M}$ to $\mathbf{M}_A$ (and vice versa) is straightforward, and to get simple mathematical relations, it is advantageous to start our analysis by considering the coherency matrix $\mathbf{C}_A$ associated with $\mathbf{M}_A$ (rather than $\mathbf{C}$). From Eqs. (2) and (4), it follows that $\mathbf{C}_A$ has the general form

arXiv:2109.06877v4 [physics.optics] (2021)

$$\mathbf{C}_A = \begin{pmatrix} \sigma_0^2 & m_{00}(\mathbf{D}_A^T + \mathbf{P}_A^T)/4 \\ m_{00}(\mathbf{D}_A + \mathbf{P}_A)/4 & \mathbf{C}_{A3} \end{pmatrix},$$

$$\mathbf{C}_{A3} \equiv \frac{1}{4}\begin{pmatrix} 4\sigma_1^2 & im_{00}D_{A3-} & -im_{00}D_{A2-} \\ -im_{00}D_{A3-} & 4\sigma_2^2 & im_{00}D_{A1-} \\ im_{00}D_{A2-} & -im_{00}D_{A1-} & 4\sigma_3^2 \end{pmatrix},$$

$$\begin{bmatrix} \sigma_0^2 \equiv m_{00}(1 + a_1 + a_2 + \varepsilon a_3), \\ \sigma_1^2 \equiv m_{00}(1 + a_1 - a_2 - \varepsilon a_3), \\ \sigma_2^2 \equiv m_{00}(1 - a_1 + a_2 - \varepsilon a_3), \\ \sigma_2^2 \equiv m_{00}(1 - a_1 - a_2 + \varepsilon a_3), \\ \mathbf{D}_A \equiv (D_{A1}, D_{A2}, D_{A3})^T, \quad \mathbf{P}_A \equiv (P_{A1}, P_{A2}, P_{A3})^T, \\ D_{Ai-} \equiv D_{Ai} - P_{Ai}, \end{bmatrix} \quad (5)$$

which can always be parameterized as

$$\mathbf{C}_A = \begin{pmatrix} \sigma_0^2 & \mu_{01}\sigma_0\sigma_1 & \mu_{02}\sigma_0\sigma_2 & \mu_{03}\sigma_0\sigma_3 \\ \mu_{01}\sigma_0\sigma_1 & \sigma_1^2 & iv_{12}\sigma_1\sigma_2 & -iv_{13}\sigma_1\sigma_3 \\ \mu_{02}\sigma_0\sigma_2 & -iv_{12}\sigma_1\sigma_2 & \sigma_2^2 & iv_{23}\sigma_2\sigma_3 \\ \mu_{03}\sigma_0\sigma_3 & iv_{13}\sigma_1\sigma_3 & -iv_{23}\sigma_2\sigma_3 & \sigma_3^2 \end{pmatrix}, \quad (6)$$

$$[\sigma_0^2 \geq \sigma_1^2 \geq \sigma_2^2 \geq \sigma_3^2, \quad \sigma_k \geq 0],$$

where the four diagonal elements $\sigma_k^2$ ($k = 0,1,2,3$) formally play the role of respective variances, while the six real parameters $\mu_{01}, \mu_{02}, \mu_{03}, v_{23}, v_{13}, v_{12}$ can be considered as correlation coefficients. Note also that the convention $\sigma_k \geq 0$ has been taken without loss of generality, while the ordering established for the diagonal elements of $\mathbf{M}_A$ ($a_3 \leq a_2 \leq a_1$) leads to $\sigma_3 \leq \sigma_2 \leq \sigma_1 \leq \sigma_0$, and vice versa.

The explicit expression of $\mathbf{M}_A$ in terms of the statistical parameters is

$$\mathbf{M}_A = \begin{pmatrix} \sigma_0^2 + \sigma_1^2 & 2\sigma_0\sigma_1\mu_{01} & 2\sigma_0\sigma_2\mu_{02} & 2\sigma_0\sigma_3\mu_{03} \\ +\sigma_2^2 + \sigma_3^2 & +2\sigma_2\sigma_3v_{23} & +2\sigma_1\sigma_3v_{13} & +2\sigma_1\sigma_2v_{12} \\ 2\sigma_0\sigma_1\mu_{01} & \sigma_0^2 + \sigma_1^2 & 0 & 0 \\ -2\sigma_2\sigma_3v_{23} & -\sigma_2^2 - \sigma_3^2 & & \\ 2\sigma_0\sigma_2\mu_{02} & 0 & \sigma_0^2 - \sigma_1^2 & 0 \\ -2\sigma_1\sigma_3v_{13} & & +\sigma_2^2 - \sigma_3^2 & \\ 2\sigma_0\sigma_3\mu_{03} & 0 & 0 & \sigma_0^2 - \sigma_1^2 \\ -2\sigma_1\sigma_2v_{12} & & & -\sigma_2^2 + \sigma_3^2 \end{pmatrix} \quad (7)$$

A detailed view of the structure of the intrinsic statistical information supported by $\mathbf{M}_A$ (hence, by $\mathbf{C}_A$) is achieved through the factorization

$$\mathbf{C}_A = \mathbf{\Sigma}\mathbf{\Gamma}\mathbf{\Sigma}, \quad (8.a)$$

where

$$\mathbf{\Sigma} \equiv 2\,\mathrm{diag}(\sigma_0, \sigma_1, \sigma_2, \sigma_3),$$

$$\mathbf{\Gamma} \equiv \frac{1}{4}\begin{pmatrix} 1 & \mu_{01} & \mu_{02} & \mu_{03} \\ \mu_{01} & 1 & iv_{12} & -iv_{13} \\ \mu_{02} & -iv_{12} & 1 & iv_{23} \\ \mu_{03} & iv_{13} & -iv_{23} & 1 \end{pmatrix}. \quad (8.b)$$

The respective coefficients 2 and $1/4$ in the definitions of $\mathbf{\Sigma}$ and $\mathbf{\Gamma}$ have been chosen so as to ensure that, besides $\mathbf{\Gamma}$ is proportional to a correlation matrix, it also has the form of a particular type of coherency matrix that, from a formal point of view, has an associated normalized Mueller matrix $\hat{\mathbf{M}}_\Gamma$ (i.e., the MIC of $\hat{\mathbf{M}}_\Gamma$ is equal to 1, which simplifies certain analyses). The Mueller matrix $\hat{\mathbf{M}}_\Gamma$, which obviously is different from $\mathbf{M}_A$, is an abstract construction that will be analyzed in Section V.

When $\sigma_3 = 0$ the correlation parameters $\mu_{03}$, $v_{23}$ and $v_{13}$ are undetermined and, without loss of generality, they can be taken as zero-valued $\mu_{03} = v_{23} = v_{13} = 0$; likewise, when $\sigma_3 = \sigma_2 = 0$ the convention that all but $\mu_{01}$ correlation parameters are zero is taken.

Therefore, since the number of zero-valued diagonal elements of matrix $\mathbf{\Sigma}$ determines the corresponding *active submatrix* of $\mathbf{\Gamma}$, the *intrinsic correlation matrix* $\mathbf{\Gamma}$ is defined as follows: when $\sigma_3 > 0$, then $\mathbf{\Gamma}$ has the form as in Eq. (8.b) (i.e., $\det \mathbf{\Sigma} > 0$, in which case, wherever appropriate, $\mathbf{\Gamma}$ is denoted as $\mathbf{\Gamma}_4$), and for the remaining cases $\mathbf{\Gamma}$ is defined as

$$\sigma_3 = 0, \sigma_2 > 0 \rightarrow \mathbf{\Gamma} \equiv \mathbf{\Gamma}_3 \equiv \frac{1}{4}\begin{pmatrix} 1 & \mu_{01} & \mu_{02} & 0 \\ \mu_{01} & 1 & iv_{12} & 0 \\ \mu_{02} & -iv_{12} & 1 & 0 \\ 0 & 0 & 0 & 0 \end{pmatrix},$$

$$\sigma_2 = 0, \sigma_1 > 0 \rightarrow \mathbf{\Gamma} \equiv \mathbf{\Gamma}_2 \equiv \frac{1}{4}\begin{pmatrix} 1 & \mu_{01} & 0 & 0 \\ \mu_{01} & 1 & 0 & 0 \\ 0 & 0 & 0 & 0 \\ 0 & 0 & 0 & 0 \end{pmatrix}, \quad (9)$$

$$\sigma_1 = 0, \sigma_0 > 0 \rightarrow \mathbf{\Gamma} \equiv \mathbf{\Gamma}_1 \equiv \frac{1}{4}\mathrm{diag}(1.0.0.0).$$

Apart from avoiding the use of undetermined extra parameters, the above convention for the definition of $\mathbf{\Gamma}$ has the key virtue that it ensures the fulfillment of the equality $\mathrm{rank}\,\mathbf{\Gamma} = \mathrm{rank}\,\mathbf{C}_A$ (recall that $\mathrm{rank}\,\mathbf{C}_A = \mathrm{rank}\,\mathbf{C}$ is always satisfied), and therefore the number of nonzero eigenvalues $\mathbf{C}_A$ (which coincides with that of $\mathbf{C}$) is equal to that of $\mathbf{\Gamma}$. Among other aspects, the physical significance of the integer parameter $r \equiv \mathrm{rank}\,\mathbf{C}$ is derived from the fact that, as discussed in Section III, $r$ is precisely the minimal number of independent parallel (incoherent) components of $\mathbf{M}$ [37,38].

From the mathematical expression $\mathbf{C}_A$ in terms of $\mathbf{M}_A$, it follows that the diagonal elements $c_k = \sigma_k^2$ of $\mathbf{C}_A$ are given by the following nonnegative quantities in terms of the diagonal elements of $\mathbf{M}_A$

$$\begin{aligned}
\sigma_0^2 &= \operatorname{tr}(\mathbf{M}_A \mathbf{M}_{Rd0}) = m_{00}(1+a_1+a_2+\varepsilon a_3)/4 \\
\sigma_1^2 &= \operatorname{tr}(\mathbf{M}_A \mathbf{M}_{Rd1}) = m_{00}(1+a_1-a_2-\varepsilon a_3)/4 \\
\sigma_2^2 &= \operatorname{tr}(\mathbf{M}_A \mathbf{M}_{Rd2}) = m_{00}(1-a_1+a_2-\varepsilon a_3)/4 \\
\sigma_3^2 &= \operatorname{tr}(\mathbf{M}_A \mathbf{M}_{Rd3}) = m_{00}(1-a_1-a_2+\varepsilon a_3)/4
\end{aligned} \quad (10)$$

where $\sigma_k^2$ have been conveniently expressed as the traces of the products of $\mathbf{M}_A$ by the respective diagonal retarders $\mathbf{M}_{Rdk}$ defined as

$$\begin{aligned}
\mathbf{M}_{Rd0} &\equiv \operatorname{diag}(1,1,1,1), \quad \mathbf{M}_{Rd1} \equiv \operatorname{diag}(1,1,-1,-1) \\
\mathbf{M}_{Rd2} &\equiv \operatorname{diag}(1,-1,1,-1), \quad \mathbf{M}_{Rd3} \equiv \operatorname{diag}(1,-1,-1,1)
\end{aligned} \quad (11)$$

[Note that, in general, $\operatorname{tr}(\mathbf{M}_A \mathbf{M}_{Rdi}) \neq \operatorname{tr}(\mathbf{M}\mathbf{M}_{Rdi})$].

The mathematical characterization of $\boldsymbol{\Gamma}$, which has one of the forms defined in Eqs. (8.b) and (9), is determined by the following conditions, (a) the absolute values of the correlation parameters are less than 1 (directly satisfied by construction of $\boldsymbol{\Gamma}$), and (b) the eigenvalues of $\boldsymbol{\Gamma}$ are nonnegative (because of $\boldsymbol{\Gamma}$ is a particular type of coherency matrix). The eigenvalues $\gamma_i$ ($i=0,1,2,3$) of $\boldsymbol{\Gamma}$ are given by

$$\boldsymbol{\Gamma}_4 \begin{cases} \gamma_0 = (1+D_\Gamma + P_\Gamma)/4 \\ \gamma_1 = (1+D_\Gamma - P_\Gamma)/4 \\ \gamma_2 = (1-D_\Gamma + P_\Gamma)/4 \\ \gamma_3 = (1-D_\Gamma - P_\Gamma)/4 \end{cases} \begin{bmatrix} D_\Gamma \equiv |\mathbf{D}_\Gamma|, \mathbf{D}_\Gamma \equiv \dfrac{\boldsymbol{\mu}+\boldsymbol{\nu}}{2} \\ P_\Gamma \equiv |\mathbf{P}_\Gamma|, \mathbf{P}_\Gamma \equiv \dfrac{\boldsymbol{\mu}-\boldsymbol{\nu}}{2} \\ \boldsymbol{\mu} \equiv (\mu_{01},\mu_{02},\mu_{03})^T \\ \boldsymbol{\nu} \equiv (\nu_{23},\nu_{13},\nu_{12})^T \end{bmatrix}$$

$$\begin{bmatrix} \gamma_3 \leq \gamma_2 \leq \gamma_1 \leq \gamma_0 \ (D_\Gamma \geq P_\Gamma) \\ \gamma_3 \leq \gamma_1 \leq \gamma_2 \leq \gamma_0 \ (D_\Gamma \leq P_\Gamma) \end{bmatrix}$$

$$\boldsymbol{\Gamma}_3 \begin{cases} \gamma_0 = (1+2D_\Gamma)/4 \\ \gamma_1 = 1/4 \\ \gamma_2 = (1-2D_\Gamma)/4 \\ \gamma_3 = 0 \end{cases} \begin{bmatrix} D_\Gamma \equiv |\mathbf{D}_\Gamma|, \mathbf{D}_\Gamma \equiv \dfrac{\boldsymbol{\mu}+\boldsymbol{\nu}}{2} \\ P_\Gamma \equiv |\mathbf{P}_\Gamma| = D_\Gamma, \mathbf{P}_\Gamma \equiv \dfrac{\boldsymbol{\mu}-\boldsymbol{\nu}}{2} \\ \boldsymbol{\mu} \equiv (\mu_{01},\mu_{02},0)^T, \boldsymbol{\nu} \equiv (0,0,\nu_{12})^T \\ \gamma_2 \leq \gamma_1 \leq \gamma_0 \end{bmatrix} \quad (12)$$

$$\boldsymbol{\Gamma}_2 \begin{cases} \gamma_0 = (1+2D_\Gamma)/4 \\ \gamma_1 = (1-2D_\Gamma)/4 \\ \gamma_2 = \gamma_3 = 0 \end{cases} \begin{bmatrix} D_\Gamma = P_\Gamma = |\mu_{01}|/2 \\ \boldsymbol{\mu} \equiv (\mu_{01},0,0)^T, \boldsymbol{\nu} \equiv (0,0,0)^T \\ \gamma_1 \leq \gamma_0 \end{bmatrix}$$

$$\boldsymbol{\Gamma}_1 \{ \gamma_0 = 1/4, \gamma_1 = \gamma_2 = \gamma_3 = 0 \}$$

Leaving aside the trivial case of $\boldsymbol{\Gamma}_1$, for each respective $\boldsymbol{\Gamma}_i$ ($i=2,3,4$) there is a single necessary and sufficient condition for the nonnegativity of $\gamma_i$, namely:
$\boldsymbol{\Gamma}_4$: $D_\Gamma + P_\Gamma \leq 1$.
$\boldsymbol{\Gamma}_3$: $D_\Gamma \leq 1/2$, i.e., $|\boldsymbol{\psi}| \leq 1$, with $|\boldsymbol{\psi}| \equiv |(\mu_{01},\mu_{02},\nu_{12})^T|$.
$\boldsymbol{\Gamma}_2$: $D_\Gamma \leq 1/2$, i.e., $|\mu_{01}| \leq 1$.

## IV. NUMBER OF PARALLEL COMPONENTS

Parallel decompositions consist of representing a depolarizing Mueller matrix as a convex sum of Mueller matrices. The physical meaning of parallel decompositions is that the incoming electromagnetic wave splits into a set of pencils that interact, without overlapping, with a number of components that are spatially distributed in the illuminated area, and the emerging pencils are incoherently recombined into the emerging beam [39].

Any depolarizing Mueller matrix $\mathbf{M}$ can be expressed as a convex sum of, at least, a number $r \equiv \operatorname{rank}\mathbf{C}$ of pure (or nondepolarizing) Mueller matrices. The general formulation of this result is based on the so-called *arbitrary decomposition* of $\mathbf{M}$ [38].

As seen in Section II, from the convention taken for the definition of $\boldsymbol{\Gamma}$, the equality $\operatorname{rank}\mathbf{C} = \operatorname{rank}\mathbf{C}_A = \operatorname{rank}\boldsymbol{\Gamma}$ is always satisfied, and therefore the inspection of the particular forms of $\boldsymbol{\Gamma}$ in Eqs. (8.b, 9) together with the explicit expressions of their eigenvalues in Eq. (12) is sufficient to analyze the achievable values of $r$. In particular

$$2 \leq \operatorname{rank}\boldsymbol{\Gamma}_4 \leq 4, \ 2 \leq \operatorname{rank}\boldsymbol{\Gamma}_3 \leq 3, \\ 1 \leq \operatorname{rank}\boldsymbol{\Gamma}_2 \leq 2, \ \operatorname{rank}\boldsymbol{\Gamma}_1 = 1, \quad (13)$$

the different values of $r$ corresponding to the following physical situations:

$r=4$. In this case, necessarily $\sigma_3 > 0$ (i.e., $\det\boldsymbol{\Sigma} > 0$): $\boldsymbol{\Gamma} = \boldsymbol{\Gamma}_4$ with $D_\Gamma + P_\Gamma < 1$.

$r=3$. Either of the following two possibilities applies: (a) $\sigma_3 > 0$: $\boldsymbol{\Gamma} = \boldsymbol{\Gamma}_4$ with $D_\Gamma + P_\Gamma = 1$ and $0 < D_\Gamma < 1$; (b) $\sigma_3 = 0$ and $\sigma_2 > 0$: $\boldsymbol{\Gamma} = \boldsymbol{\Gamma}_3$ with $|\boldsymbol{\psi}| < 1$, $\boldsymbol{\psi} \equiv (\mu_{01},\mu_{02},\nu_{12})^T$.

$r=2$. Either of the following three possibilities holds: (a) $\sigma_3 > 0$: $\boldsymbol{\Gamma} = \boldsymbol{\Gamma}_4$ with $\boldsymbol{\nu} = \pm\boldsymbol{\mu}$ and $|\boldsymbol{\mu}| = 1$, so that the eigenvalues of $\boldsymbol{\Gamma}_4$ are given by $\gamma_0 = \gamma_1 = 1/2$, $\gamma_2 = \gamma_3 = 0$; (b) $\sigma_3 = 0$ and $\sigma_2 > 0$: $\boldsymbol{\Gamma} = \boldsymbol{\Gamma}_3$ with $|\boldsymbol{\psi}| = 1$, $|\boldsymbol{\psi}| \equiv (\mu_{01},\mu_{02},\nu_{12})^T$; (c) $\sigma_3 = \sigma_2 = 0$ and $\sigma_1 > 0$: $\boldsymbol{\Gamma} = \boldsymbol{\Gamma}_2$ and $\mu_{01}^2 < 1$.

$r=1$. Either of the following two possibilities applies: (a) $\sigma_2 = 0$ and $\sigma_1 > 0$: $\boldsymbol{\Gamma} = \boldsymbol{\Gamma}_2$ with $\mu_{01}^2 = 1$; (b) $\sigma_1 = 0$ and $\sigma_0 > 0$: $\boldsymbol{\Gamma} = \boldsymbol{\Gamma}_1$.

The above analysis shows the way in which the intrinsic variances and correlations regulate the structure of $\mathbf{M}$ in terms of the minimal number $r$ of pure parallel components of $\mathbf{M}$. Thus, pure (nondepolarizing) systems correspond to the case $r=1$, which are incompatible with structures where $\sigma_2 > 0$. Depolarizing systems ($r > 1$) are characterized by the structures indicated above, depending on the number of independent components ($r = 2, 3, 4$).

Consequently, the statistical formulation in Eq. (8), together with the definition of $\boldsymbol{\Gamma}$, provides, in a simple way, the general structure of $r$-component depolarizing Mueller matrices ($2 \leq r \leq 4$). Note that the approach presented in [40], based on the enpolarizing ellipsoid, provides an alternative formulation of the structure of Mueller matrices with $r = 2$.

## V. THE CORRELATION MUELLER MATRIX

As seen in Section III, given a correlation matrix $\boldsymbol{\Gamma}$, from a formal point of view it has an associated *correlation Mueller matrix* (CMM), $\hat{\mathbf{M}}_\Gamma$, whose elements are obtained from Eq. (2) by replacing $\mathbf{C}$ by $\boldsymbol{\Gamma}$. Thus, $\hat{\mathbf{M}}_\Gamma$ is likewise associated with $\mathbf{M}_A$ (and with $\mathbf{M}$, through $\mathbf{M}_A$) and has the peculiar and simple form

$$\hat{\mathbf{M}}_\Gamma = \begin{pmatrix} 1 & \mathbf{D}_\Gamma^T \\ \mathbf{P}_\Gamma & \mathbf{0} \end{pmatrix}, \qquad (14)$$

where vectors $\mathbf{D}_\Gamma = (\boldsymbol{\mu} + \boldsymbol{\nu})/2$ and $\mathbf{P}_\Gamma = (\boldsymbol{\mu} - \boldsymbol{\nu})/2$ are, respectively, the diattenuation and polarizance vectors of $\hat{\mathbf{M}}_\Gamma$.

Despite the obvious fact that $\hat{\mathbf{M}}_\Gamma \ne \hat{\mathbf{M}}_A$ and that arrow Mueller matrices with different associated $\boldsymbol{\Sigma}$ can share a common associated $\hat{\mathbf{M}}_\Gamma$, $\hat{\mathbf{M}}_\Gamma$ encodes, uniquely, all information on the two *intrinsic correlation vectors*, $\boldsymbol{\mu}$ and $\boldsymbol{\nu}$, while the remaining information on the four $\sigma_k$ is contained in matrix $\boldsymbol{\Sigma}$ in a decoupled and exclusive manner.

Since $P_S(\hat{\mathbf{M}}_\Gamma) = 0$, then $P_\Delta^2(\hat{\mathbf{M}}_\Gamma) = (\mathbf{D}_\Gamma^2 + \mathbf{P}_\Gamma^2)/3$, showing that the contributions to the degree of polarimetric purity of $\hat{\mathbf{M}}_\Gamma$ come only from the enpolarizing components, which in turn implies that $P_\Delta(\hat{\mathbf{M}}_\Gamma) \le 1/\sqrt{3}$, whose maximum $P_\Delta(\hat{\mathbf{M}}_\Gamma) = 1/\sqrt{3}$ is achieved when either (a) $D_\Gamma = 1$ and $P_\Gamma = 0$, or (b) $D_\Gamma = 0$ and $P_\Gamma = 1$ (because of condition $D_\Gamma + P_\Gamma \le 1$).

Thus, the relation between the polarimetric purities $P_\Delta(\mathbf{M})$ and $P_\Delta(\hat{\mathbf{M}}_\Gamma)$ is not biunivocal, but is critically mediated by the values of $\sigma_k$. This is evidenced, for instance, from the fact that $P_\Delta(\mathbf{M}_\Gamma) = 0$ may correspond either to $P_\Delta(\mathbf{M}) = 1$ or to $P_\Delta(\hat{\mathbf{M}}_\Gamma) = 0$ depending on the particular form of $\mathbf{M}$.

It is remarkable that the degree of polarizance $P_P(\mathbf{M})$ of $\mathbf{M}$, defined as $P_P \equiv \sqrt{(P^2 + D^2)/2}$ [which always satisfies $P_P(\mathbf{M}) = P_P(\mathbf{M}_A)$] [29], is zero if and only if $P_P(\hat{\mathbf{M}}_\Gamma) = 0$, showing how the information on $P_P(\mathbf{M})$ and $P_S(\mathbf{M})$ encoded in $\mathbf{M}_A$ is decoupled in a peculiar manner through the transformation $\mathbf{C}_A = \boldsymbol{\Sigma}\boldsymbol{\Gamma}\boldsymbol{\Sigma}$ in Eq. (8).

## VI. CHARACTERIZATION THEOREM

As a consequence of the powerful decoupling features of the statistical approach formulated above, the general characterization of Mueller matrices can be stated as follows through particularly simple conditions:

Given a 4×4 real matrix $\mathbf{M}$, it is a Mueller matrix if and only if the following conditions hold:
*Variance conditions.* $\mathrm{tr}(\mathbf{M}_A \mathbf{M}_{Rdk}) \ge 0$ [$k = 0,1,2,3$, see Eqs. (10) and (11)].
*Correlation condition.* $D_\Gamma + P_\Gamma \le 1$, $D_\Gamma$ and $P_\Gamma$ being the diattenuation and polarizance of the correlation Mueller matrix $\hat{\mathbf{M}}_\Gamma$ associated with $\mathbf{M}$. This single correlation condition can also be expressed as $|\boldsymbol{\mu} + \boldsymbol{\nu}| + |\boldsymbol{\mu} - \boldsymbol{\nu}| \le 2$.
*Passivity condition.* $m_{00}(1+Q) \le 1$, with $Q \equiv \max(D,P)$ [10,29].

Observe that the four covariance conditions established by Cloude through the nonnegativity of the eigenvalues of $\mathbf{C}(\mathbf{M})$ [2] are fully equivalent to the set of five conditions composed of the combination of the four (trivial) variance inequalities and the correlation condition. Even though the above characterization involves five conditions (plus the passivity one) instead of the four Cloude's conditions (plus the passivity one), the advantage of the new approach is that the characterization is made in terms of intrinsic and physically meaningful properties that are directly expressed in terms of the Mueller matrices $\mathbf{M}_A$ and $\hat{\mathbf{M}}_\Gamma$ associated with $\mathbf{M}$.

The determination of the intrinsic polarimetric information provided by $\mathbf{M}$, which is contained in $\boldsymbol{\Sigma}$, $\boldsymbol{\mu}$ and $\boldsymbol{\nu}$, can be performed straightforwardly through the transformations $\mathbf{M} \to \mathbf{M}_A \to \mathbf{C}_A \to \boldsymbol{\Sigma},\boldsymbol{\mu},\boldsymbol{\nu}$, while the retardance information is decoupled and uniquely provided by $\mathbf{M}_{RI}$ and $\mathbf{M}_{RO}$. (Note that, despite the fact that the so-called *birefringence anisotropy coefficients* [36,41] of $\mathbf{M}_A$ are zero, the possible retardance properties exhibited by $\mathbf{M}_A$ require further analysis, which will be performed in a future work).

Conversely, the complete set of Mueller matrices can be synthesized though the following procedure:
1) Take four arbitrary nonnegative parameters $1 \ge \sigma_0 \ge \sigma_1 \ge \sigma_2 \ge \sigma_3 \ge 0$ and build matrix $\boldsymbol{\Sigma} \equiv 2\,\mathrm{diag}(\sigma_0,\sigma_1,\sigma_2,\sigma_3)$.
2) Take a pair of arbitrary three-component real vectors $\boldsymbol{\mu}$ and $\boldsymbol{\nu}$ that satisfy the inequalities $|\boldsymbol{\mu}| \le 1$, $|\boldsymbol{\nu}| \le 1$, and $|\boldsymbol{\mu}+\boldsymbol{\nu}|/2 + |\boldsymbol{\mu}-\boldsymbol{\nu}|/2 \le 1$.
3) Calculate the corresponding arrow Mueller matrix $\mathbf{M}_A(\boldsymbol{\Sigma},\boldsymbol{\mu},\boldsymbol{\nu})$ by means of Eq. (7).
4) Take a physically realizable MIC, $m_{00}$, that satisfies $m_{00} \le 1/(1+Q)$ [$Q \equiv \min(D,P)$], so that $\mathbf{M}_A = m_{00} \hat{\mathbf{M}}_A$.
5) Take (freely) two orthogonal Mueller matrices $\mathbf{M}_{RI}$ and $\mathbf{M}_{RO}$ of the respective entrance and exit retarders.
6) Build the corresponding Mueller matrix through the dual retarder transformation $\mathbf{M} = \mathbf{M}_{RO} \mathbf{M}_A \mathbf{M}_{RI}$.

## VII. STRUCTURE AND SIGNIFICANCE OF THE POLARIMETRIC INFORMATION SUPPORTED BY A MUELLER MATRIX

Once the statistical structure of the coherency matrix $\mathbf{C}_A$ associated with a given Mueller matrix $\mathbf{M}$ (through its arrow form $\mathbf{M}_A$) has been determined by means of specific conditions on matrix $\boldsymbol{\Sigma}$ and on the two intrinsic correlation enpolarizing vectors $\boldsymbol{\mu}$ and $\boldsymbol{\nu}$, it is worth considering the statistical structure of the coherency matrix $\mathbf{C}$ associated with a general Mueller matrix $\mathbf{M}$, which can always be expressed as $\mathbf{C} = \boldsymbol{\Sigma}\boldsymbol{\Omega}\boldsymbol{\Sigma}$, with

$$\boldsymbol{\Omega} = \frac{1}{4}\begin{pmatrix} 1 & \mu_{01} + i\eta_{01} & \mu_{02} - i\eta_{02} & \mu_{03} + i\eta_{03} \\ \mu_{01} - i\eta_{01} & 1 & \tau_{12} + i\nu_{12} & \tau_{13} - i\nu_{13} \\ \mu_{02} + i\eta_{02} & \tau_{12} - i\nu_{12} & 1 & \tau_{23} + i\nu_{23} \\ \mu_{03} - i\eta_{03} & \tau_{13} + i\nu_{13} & \tau_{23} - i\nu_{23} & 1 \end{pmatrix}, \quad (15)$$

$$\left[ \sigma_k, \mu_{0k}, \nu_{lk}, \eta_{0k}, \tau_{lk} \in \mathbb{R},\ \sigma_k \ge 0 \atop l,k = 1,2,3\ \ l < k \right],$$

where, for this general case, the equivalent standard deviations $\sigma_k$ are not subject to the ordering restriction

$\sigma_0 \geq \sigma_1 \geq \sigma_2 \geq \sigma_3$ taken for the arrow form, and two additional constitutive real vectors, the *correlation retardance vectors* $\boldsymbol{\eta} \equiv (\eta_{01}, \eta_{02}, \eta_{03})^T$ and $\boldsymbol{\tau} \equiv (\tau_{23}, \tau_{13}, \tau_{12})^T$, are identified. The latter vanish for $\mathbf{M}_A$, and as we will see below, up to the modulation produced by $\sigma_k$, vectors $\boldsymbol{\eta}$ and $\boldsymbol{\tau}$ encode the specific information on the entrance and exit retardance properties.

The key role played by $\sigma_k$ and the correlation vectors $\boldsymbol{\mu}, \boldsymbol{\nu}, \boldsymbol{\eta}$, and $\boldsymbol{\tau}$ in the information structure of $\mathbf{M}$, becomes explicit when $\mathbf{M}$ is expressed in terms of them

$$\mathbf{M} = \begin{pmatrix} \sigma_0^2 + \sigma_1^2 + \sigma_2^2 + \sigma_3^2 & 2\sigma_0\sigma_1\mu_{01} + 2\sigma_2\sigma_3\nu_{23} & 2\sigma_0\sigma_2\mu_{02} + 2\sigma_1\sigma_3\nu_{13} & 2\sigma_0\sigma_3\mu_{03} + 2\sigma_1\sigma_2\nu_{12} \\ 2\sigma_0\sigma_1\mu_{01} - 2\sigma_2\sigma_3\nu_{23} & \sigma_0^2 + \sigma_1^2 - \sigma_2^2 - \sigma_3^2 & -2\sigma_0\sigma_3\eta_{03} + 2\sigma_1\sigma_2\tau_{12} & -2\sigma_0\sigma_2\eta_{02} + 2\sigma_1\sigma_3\tau_{13} \\ 2\sigma_0\sigma_2\mu_{02} - 2\sigma_1\sigma_3\nu_{13} & 2\sigma_0\sigma_3\eta_{03} + 2\sigma_1\sigma_2\tau_{12} & \sigma_0^2 - \sigma_1^2 + \sigma_2^2 - \sigma_3^2 & -2\sigma_0\sigma_1\eta_{01} + 2\sigma_2\sigma_3\tau_{23} \\ 2\sigma_0\sigma_3\mu_{03} - 2\sigma_1\sigma_2\nu_{12} & 2\sigma_0\sigma_2\eta_{02} + 2\sigma_1\sigma_3\tau_{13} & 2\sigma_0\sigma_1\eta_{01} + 2\sigma_2\sigma_3\tau_{23} & \varepsilon\sigma_0^2 - \varepsilon\sigma_1^2 - \varepsilon\sigma_2^2 + \varepsilon\sigma_3^2 \end{pmatrix} \quad (16)$$

Despite the notation used, it should be stressed that, except when $\mathbf{M} = \mathbf{M}_A$, $\boldsymbol{\Sigma}$ and the correlation enpolarizing vectors of $\mathbf{M}$ are different from those of $\mathbf{M}_A$. Also, as with matrix $\boldsymbol{\Gamma}$, and without loss of generality, it is adopted the convention that when one or more $\sigma_k$ are zero-valued the corresponding $k$ rows and columns of $\boldsymbol{\Omega}$ are zero).

Observe that the diattenuation and polarizance vectors of $\mathbf{M}$ only depend on $\sigma_k$ and vectors $\boldsymbol{\mu}$ and $\boldsymbol{\nu}$ (of $\mathbf{M}$), and that $\boldsymbol{\mu}$ and $\boldsymbol{\nu}$ (of $\mathbf{M}$) do not take place in the submatrix $\mathbf{m}$ of $\mathbf{M}$.

To go deeper into the analysis of the information structure of $\mathbf{M}$, observe that it can be expressed as follows in terms of five constitutive vectors

$$\mathbf{M} \equiv m_{00} \begin{pmatrix} 1 & D_1 & D_2 & D_3 \\ P_1 & k_1 & r_3 & r_2 \\ P_2 & q_3 & k_2 & r_1 \\ P_3 & q_2 & q_1 & k_3 \end{pmatrix}, \quad (17)$$

$$\mathbf{k} \equiv \frac{1}{\sqrt{3}}\begin{pmatrix}k_1\\k_2\\k_3\end{pmatrix}, \mathbf{D} \equiv \begin{pmatrix}D_1\\D_2\\D_3\end{pmatrix}, \mathbf{P} \equiv \begin{pmatrix}P_1\\P_2\\P_3\end{pmatrix}, \mathbf{r} \equiv \begin{pmatrix}r_1\\r_2\\r_3\end{pmatrix}, \mathbf{q} \equiv \begin{pmatrix}q_1\\q_2\\q_3\end{pmatrix}$$

so that the diagonal elements of $\mathbf{M}$, depend exclusively on the variances $\sigma_k^2$

$$\mathbf{k} = \frac{1}{\sqrt{3}\, m_{00}}\begin{pmatrix}\sigma_0^2 + \sigma_1^2 - \sigma_2^2 - \sigma_3^2 \\ \sigma_0^2 - \sigma_1^2 + \sigma_2^2 - \sigma_3^2 \\ \sigma_0^2 - \sigma_1^2 - \sigma_2^2 + \sigma_3^2\end{pmatrix}, \quad (18)$$

$$[m_{00} = \sigma_0^2 + \sigma_1^2 + \sigma_2^2 + \sigma_3^2],$$

while the other four constitutive vectors can appropriately be transformed into another set where the different types of correlations are decoupled each other

$$\mathbf{P}_+ \equiv \frac{\mathbf{D}+\mathbf{P}}{2} = \frac{2}{m_{00}}\bar{\boldsymbol{\Sigma}}_0\bar{\boldsymbol{\Sigma}}_1\boldsymbol{\mu}, \quad \mathbf{P}_- \equiv \frac{\mathbf{D}-\mathbf{P}}{2} = \frac{2}{m_{00}}\bar{\boldsymbol{\Sigma}}_2\bar{\boldsymbol{\Sigma}}_3\boldsymbol{\nu},$$

$$\mathbf{R}_+ \equiv \frac{\mathbf{q}+\mathbf{r}}{2} = \frac{2}{m_{00}}\bar{\boldsymbol{\Sigma}}_2\bar{\boldsymbol{\Sigma}}_3\boldsymbol{\tau}, \quad \mathbf{R}_- \equiv \frac{\mathbf{q}-\mathbf{r}}{2} = \frac{2}{m_{00}}\bar{\boldsymbol{\Sigma}}_0\bar{\boldsymbol{\Sigma}}_1\boldsymbol{\eta}, \quad (19)$$

$$\begin{bmatrix}\bar{\boldsymbol{\Sigma}}_0 \equiv \mathrm{diag}(\sigma_0,\sigma_0,\sigma_0), \bar{\boldsymbol{\Sigma}}_1 \equiv \mathrm{diag}(\sigma_1,\sigma_2,\sigma_3),\\ \bar{\boldsymbol{\Sigma}}_2 \equiv \mathrm{diag}(\sigma_2,\sigma_3,\sigma_1), \bar{\boldsymbol{\Sigma}}_3 \equiv \mathrm{diag}(\sigma_3,\sigma_1,\sigma_2).\end{bmatrix}$$

Thus, vectors $\mathbf{P}_+$ and $\mathbf{P}_-$, are directly and exclusively related to the correlation enpolarizing vectors $\boldsymbol{\mu}$ and $\boldsymbol{\nu}$, respectively, while vectors $\mathbf{R}_+$ and $\mathbf{R}_-$ are directly and exclusively related to the correlation retardance vectors $\boldsymbol{\eta}$ and $\boldsymbol{\tau}$, respectively. In fact, $\mathbf{R}_+ = \mathbf{R}_- = \mathbf{0}$ (or, equivalently, $\mathbf{q} = \mathbf{r} = \mathbf{0}$), if and only if $\boldsymbol{\eta} = \boldsymbol{\tau} = \mathbf{0}$, and $\mathbf{M}$ lacks enpolarizing properties (i.e., $\mathbf{P}_+ = \mathbf{P}_- = \mathbf{0}$, or, equivalently, $\mathbf{D} = \mathbf{P} = \mathbf{0}$) if and only if $\boldsymbol{\mu} = \boldsymbol{\nu} = \mathbf{0}$. In summary, all polarimetric information is structured through the four variances $\sigma_k^2$ and the four characteristic vectors $\mathbf{P}_+, \mathbf{P}_-, \mathbf{R}_+, \mathbf{R}_-$, or equivalently, through $m_{00}$ together with vectors $(\mathbf{k}, \mathbf{P}_+, \mathbf{P}_-, \mathbf{R}_+, \mathbf{R}_-)$.

## VIII CHARACTERIZATION OF NONENPOLARIZING MUELLER MATRICES

Certain polarimetric interactions are presented by *nonenpolarizing Mueller matrices*, that is, Mueller matrices $\mathbf{M}_O$ with zero diattenuation and polarizance. This occurs, for instance, for parallel compositions of retarders. In fact, any nonenpolarizing Mueller matrix is equivalent to a parallel mixture of retarders [39]. Due to the lack of polarizance and diattenuation, $\mathbf{M}_O$ has necessarily the diagonal arrow form

$$\mathbf{M}_A(\mathbf{M}_O) = m_{00}\, \mathrm{diag}(1, a_1, a_2, \varepsilon a_3), \quad (20)$$

and consequently, $\mathbf{C}_A(\mathbf{M}_O) = \mathrm{diag}(\sigma_0^2, \sigma_1^2, \sigma_2^2, \sigma_3^2)$, where $\sigma_k^2$ just coincide with the eigenvalues of $\mathbf{C}_A(\mathbf{M}_O)$. Therefore, $\hat{\mathbf{M}}_\Gamma = \mathrm{diag}(1,0,0,0)$, and, in terms of the statistical parameters, the covariance conditions become the trivial exigency that $\sigma_k^2$ are nonnegative, together with $D_\Gamma = P_\Gamma = 0$ (i.e., all intrinsic correlation coefficients are zero).

## IX CHARACTERIZATION OF SYMMETRIC MUELLER MATRICES

Many physical situations encountered in polarimetry involve Mueller matrices that are symmetric. By considering the arrow form of a symmetric Mueller matrix $\mathbf{M}_S$, it turns out that necessarily the entrance and exit retarders are inverse each other ($\mathbf{M}_{RO} = \mathbf{M}_{RI}^T$), and furthermore the arrow form $\mathbf{M}_{AS}$ of $\mathbf{M}_S$ is also a symmetric matrix, so that $\mathbf{M}_S = \mathbf{M}_R^T \mathbf{M}_{AS} \mathbf{M}_R$. Observe also that the coherency matrix $\mathbf{C}_A(\mathbf{M}_S)$ has itself a symmetric form (i.e., $\boldsymbol{\nu} = \mathbf{0}$), and consequently the set of covariance conditions reduces to the trivial inequalities $\sigma_k^2 \geq 0$ together with $D_\Gamma = P_\Gamma \leq 1/2$.

## X. CONCLUSION

Any macroscopic linear polarimetric interaction, represented by the corresponding Mueller matrix **M**, necessarily results from an average a myriad of elementary nondepolarizing interactions, which shows that **M** has an essentially statistical nature. The mathematical structure of **M** makes the sixteen elements of **M** to be related, in an intricate and coupled manner, to the phenomenological quantities that provide a direct interpretation of the polarimetric information supported by **M**.

To achieve a simple general characterization and physical interpretation of such information, the approach presented is based on the combination of the arrow form $\mathbf{M}_A$ of **M** (defined and studied in previous works [29]) and the factorization of the coherency matrix $\mathbf{C}_A$ associated with $\mathbf{M}_A$ in terms of an intrinsic correlation matrix $\mathbf{\Gamma}$ that is pre- and post-multiplied by the intrinsic diagonal matrix $\mathbf{\Sigma}$. The definition of $\mathbf{\Gamma}$ is performed in such a way that the extra correlation variables that do not take place in the problem are avoided, which simplifies the mathematical formulation. Furthermore, this ensures that the ranks of matrices $\mathbf{C}_A$ and $\mathbf{\Gamma}$ are equal, which is a key feature for obtaining the desired results of decoupling and interpreting the relevant physical information.

Thus, $\mathbf{\Gamma}$ can formally be considered as a coherency matrix, and it is found that its associated correlation Mueller matrix $\hat{\mathbf{M}}_\Gamma$ has an extremely simple structure that depends exclusively on the essential enpolarizing properties and allows for a simple general characterization of Mueller matrices that is formulated through (a) the nonnegativity of four equivalent variances $\sigma_k^2$ (expressed directly in terms of properties of **M**, without the necessity of using auxiliary Hermitian matrices); (b) the inequality, $D_\Gamma + P_\Gamma \leq 1$, which involves only the absolute values of the pair of intrinsic correlation enpolarizing vectors, and (c) the well known passivity condition.

The general structure of information is then analyzed in terms of five constitutive vectors and their relations to the basic physical properties.

The total decoupling of retardance properties achieved through the arrow decomposition, which was proposed in previous works, is reaffirmed in a natural manner. In addition, the essential entanglement among enpolarizing and depolarizing properties appears as a consequence of the critical effect of the equivalent variances.

Given a Mueller matrix, the obtainment of $\mathbf{\Sigma}$ (whose elements are the equivalent variances), the intrinsic correlation enpolarizing vectors (constitutive of $\mathbf{\Gamma}$), as well as the equivalent entrance and exit retarders is straightforward. Conversely, any Mueller matrix can be synthesized by applying simple criteria to the choice of the intrinsic parameters and the entrance and exit retarders.

The characterization of the particular (but important in practice) types of nonenpolarizing and symmetric Mueller matrices is performed straightforwardly by means of the general characterization theorem introduced.

**REFERENCES**

1. Gil, J. J. and Bernabéu, E. A depolarization criterion in Mueller matrices. Optica Acta **32**, 259–261 (1985).
2. S.R. Cloude, "Group theory and polarization algebra", Optik **75**, 26-36 (1986).
3. K. Kim, L. Mandel, E. Wolf, Relationship between Jones and Mueller matrices for random media, J. Opt. Soc. Amer. A **4** 433–437 (1987).
4. C. V. M. van der Mee, "An eigenvalue criterion for matrices transforming Stokes parameters," J. Math. Phys. **34** 5072-5088 (1993).
5. R. Sridhar and R. Simon, "Normal form for Mueller matrices in polarization optics," J. Mod Opt. **41**, 1903-1915 (1994).
6. Y. Bolshakov, C. V. M. van der Mee, A. C. M. Ran, B. Reichstein, and L. Rodman, "Polar decompositions in finite dimensional indefinite scalar product spaces: special cases and applications," in *Operator Theory: Advances and Applications, I.* Gohberg, ed. (Birkhäuser Verlag, 1996), Vol. 87, pp. 61–94.
7. Y. Bolshakov, C. V. M. van der Mee, A. C. M. Ran, B. Reichstein, and L. Rodman, "Errata for: Polar decompositions in finite dimensional indefinite scalar product spaces: special cases and applications," Integral Equation Oper. Theory **27**) 497–501 (1997.
8. A.V. Gopala Rao, K.S. Mallesh, and Sudha, "On the algebraic characterization of a Mueller matrix in polarization optics. I. Identifying a Mueller matrix from its *N* matrix", J. Mod. Opt. **45** 955-987 (1998).
9. A.V. Gopala Rao, K.S. Mallesh, and Sudha, "On the algebraic characterization of a Mueller matrix in polarization optics. II. Necessary and sufficient conditions for Jones derived Mueller matrices", J. Mod. Opt. **45** 989-999 (1998).
10. J. J. Gil, "Characteristic properties of Mueller matrices" J. Opt. Soc. Am. A **17** 328-334 (2000).
11. T. Tudor, "Generalized observables in polarization optics," J. Phys. A **36** 9577-9590 (2003).
12. R. Ossikovski, "Analysis of depolarizing Mueller matrices through a symmetric decomposition", J. Opt. Soc. Am. A **26**, 1109-1118 (2009).
13. R. Ossikovski, "Canonical forms of depolarizing Mueller matrices", J. Opt. Soc. Am. A **27**, 123-130 (2010).
14. B. N. Simon, S. Simon, N. Mukunda, F. Gori, M. Santarsiero, R. Borghi, and R. Simon, "A complete characterization of pre-Mueller and Mueller matrices in polarization optics", J. Opt. Soc. Am. A **27**, 188-199 (2010).
15. S. R. Cloude, "Depolarization synthesis: understanding the optics of Mueller matrix depolarization", J. Opt. Soc. Am A **30**, 691-700 (2013)
16. J. J. Gil, I. San José, "Explicit algebraic characterization of Mueller matrices", Opt. Lett. **13**, 4041-4044 (2014).
17. R. Ossikovski, J. J. Gil, "Basic properties and classification of Mueller matrices derived from their statistical definition", J. Opt. Soc. Am. A **34**, 1727-1737 (2017).
18. J. J. Gil, "Components of purity of a Mueller matrix", J. Opt. Soc. Am. A **28**, 1578-1585 (2011).
19. N. G. Parke, III, *Matrix optics*. Ph.D. dissertation (MIT, Cambridge, Mass., 1948).
20. R. Ossikovski, K. Hingerl, "General formalism for partial spatial coherence in reflection Mueller matrix polarimetry", Opt. Lett. **41**, 4044-4047 (2016)
21. K. Hingerl, R. Ossikovski, "General approach for modeling partial coherence in spectroscopic Mueller matrix polarimetry", Opt. Lett. **41**, 219-222 (2016).

22. E. Kuntman, M. A. Kuntman, J. Sancho-Parramon, O. Arteaga, "Formalism of optical coherence and polarization based on material media states", Phys. Rev. A **95**, 063819 (2017).
23. I. San José, J. J. Gil, "Coherency vector formalism for polarimetric transformations", Opt. Commun. 475, 126230 (2020).
24. T. Setälä, A. Shevchenko, M. Kaivola, A. T. Friberg, "Polarization time and length for random optical beams", Phys. Rev. A **78**, 033817 (2008).
25. A. Shevchenko, T. Setälä, A. M. Kaivola, A. T. Friberg "Characterization of polarization fluctuations in random electromagnetic beams", New J. Phys. **11**, 073004 (2009)
26. A. Shevchenko, M. Roussey, A. T. Friberg, T. Setälä "Polarization time of unpolarized light", Optica **4** 64-70 (2017).
27. B. A. Robson, *The Theory of Polarization Phenomena*. (Clarendon, 1974).
28. Z-F Xing, "On the deterministic and non-deterministic Mueller matrix," J. Mod. Opt. **39**, 461-484 (1992).
29. I. San José, J. J. Gil, "characterization of passivity in Mueller matrices", J. Opt. Soc. Am. A **37**, 199-208 (2020).
30. J. J. Gil, "Transmittance constraints in serial decompositions of depolarizing Mueller matrices. The arrow form of a Mueller matrix", J. Opt. Soc. Am. A **30**, 701-707 (2013).
31. J. J. Gil, "Structure of polarimetric purity of a Mueller matrix and sources of depolarization", Opt. Commun. **368**, 165–173 (2016).
32. J. J. Gil, "Review on Mueller matrix algebra for the analysis of polarimetric measurements", J. Appl. Remote Sens. **8**, 081599-37 (2014).
33. J. J. Gil, "Invariant quantities of a Mueller matrix under rotation and retarder transformations," J. Opt. Soc. Am. A **33**, 52-58 (2016).
34. I. San José, J. J. Gil, "Invariant indices of polarimetric purity: generalized indices of purity for n× n covariance matrices", Opt. Commun. **284**, 38-47 (2011).
35. J. J. Gil, E. Bernabéu, "Depolarization and polarization indices of an optical system," Opt. Acta **33**, 185-189 (1986).
36. J. J. Gil and R. Ossikovski, *Polarized Light and the Mueller Matrix Approach* (CRC Press, 2016).
37. J. J. Gil, I. San José, "Polarimetric subtraction of Mueller matrices", J. Opt. Soc. Am. A **30**, 1078-1088 (2013).
38. J. J. Gil, I. San José, "Arbitrary decomposition of Mueller matrices," Opt. Letters 44, 5715-5718 (2019).
39. J. J. Gil, "Polarimetric characterization of light and media," Eur. Phys. J. Appl. Phys. **40**, 1-47 (2007).
40. J. J. Gil, I. San José, "Universal Synthesizer of Mueller Matrices Based on the Symmetry Properties of the Enpolarizing Ellipsoid", Symmetry, **13**, 983 (2021).
41. O. Arteaga, E. García-Caurel, R. Ossikovski, "Anisotropy coefficients of a Mueller matrix," J. Opt. Soc. Am. A **28**, 548-553 (2011).